\documentclass{ws-p8-50x6-00}
\newcommand{\lsim}{\raisebox{-4pt}{$\,\stackrel{\textstyle
                                                         <}{\sim}\,$}}

\newcommand{\da}{{distribution amplitude}}
\def\qb{\overline{Q}}
\def\qqb{\overline{Q}\hspace{1pt}^2}
\def\als{\alpha_s}
\def\mev{\,{\rm MeV}}
\def\gev{\,{\rm GeV}}

\begin{document}

\title{The transition of virtual photons into pseudoscalar mesons}

\author{P.\ Kroll}

\address{Fachbereich Physik, Universit\"at Wuppertal, 42097 Wuppertal,
Germany\\ 
E-mail: kroll@theorie.physik.uni-wuppertal.de }

\maketitle

\abstracts{
The possibility to constrain the meson \da{} from $\gamma^*\gamma^*\to
\pi, \eta, \eta'$ transitions is investigated. It is shown that for a
large range in the two photon virtualities the transition form factors
are essentially independent of the \da s. This in turn entails 
parameter-free predictions of QCD.}

Since the advent of the CLEO measurement \cite{cleo98} of the $\gamma^*
- P$ transition form factor ($P=\pi, \eta, \eta'$) for quasi-real
photons many papers appeared that have been devoted to the theoretical
analysis of these form factors. It became evident from these analyses
of the CLEO data that the \da{}s for pseudoscalar mesons are close to
the asymptotic form, $\Phi_{\rm AS}(\xi)= 3 (1-\xi^2)/2$,
where $\xi=2x-1$, and $x$ is the usual momentum fraction carried by
the quark inside the meson. This result, although not very precise, 
had a strong impact on the phenomenology of hard exclusive
reactions. Thus, for instance, earlier conjectures of large
contributions from soft physics to the pion's electromagnetic form
factor or to two-photon annihilations into pairs of pseudoscalar
mesons became substantiated. These contributions, although formally
representing power corrections to the asymptotically leading twist ones,
seem to dominate for momentum transfers of the order of a few GeV.
The meson \da{}s are also an input for the calculation of charmonium
or $B$-meson decays into pairs of pseudoscalar mesons. A good
knowledge of the \da{}s would enhance prospects of extracting
information on CP violations from the latter process.

In this talk I am going to report on a recent paper by M.\ Diehl, C.\
Vogt and myself \cite{DKV} where we investigated what information on
the \da{}s can be extracted from $\gamma^*\gamma^*\to P$ transitions
beyond that what has been obtained from the CLEO data in the real-photon
limit.

Let me begin with the discussion of the  $\gamma^*\gamma^*\to \pi$
transitions. To leading twist accuracy the
transition form factor $F_{\pi\gamma^*}$ reads \cite{agu81,bra83}  
\begin{equation}
F_{\pi\gamma^*}(\qb,\omega) = 
\frac{f_\pi}{3\sqrt{2}\, \qqb}\,
\int_{-1}^{\;1} {d} \xi\, \frac{\Phi_\pi(\xi,\mu_F)}{1-\xi^2\omega^2}\, 
\left[1 + \frac{\als(\mu_R)}{\pi}\,{\cal K}(\omega,\xi,\qb/\mu_F) 
\right] \,,
\label{fpgvirtual}
\end{equation}
where $\qqb = (Q^2 + Q'^2)/2$ and $\omega= (Q^2 - Q'^2)/(Q^2 + Q'^2)$.
$Q^2$ and $Q'^2$ denote the (space-like) virtualities of the
photons. The hard scattering kernel has been calculated to
next-to-leading oder (NLO); the expression for ${\cal K}$ can be found
in \cite{DKV,agu81,bra83}. The factorization, $\mu_F$, and 
renormalization, $\mu_R$, scales, are chosen to be equal to 
$\overline{Q}$ here. $f_\pi\approx 131 \mev$ is the pion decay
constant. The pion \da, $\Phi_\pi$, can be expanded~\cite{lep79}
upon Gegenbauer polynomials $C_n^{3/2}(\xi)$, 
$\Phi_{\pi}=\Phi_{\rm AS} \big[1+ \sum B_n (\mu_F)\,C_n^{3/2}(\xi)\big] \,$.
The Gegenbauer coefficients, $B_n$, which encode the soft physics
information required in the calculation of the form factor, evolve
with the scale $\mu_F$.
Using the expansion of the \da{}, the integrals in
(\ref{fpgvirtual}) can be worked out analytically order by order in
$n$. This results in (for details see \cite{DKV})
\begin{equation}
F_{\pi\gamma^*}(\qb,\omega) = \frac{f_{\pi}}{\sqrt{2}\: \qqb} 
 \Big[ c_0(\omega,\mu_R) 
 + \sum_{n=2,4,\ldots} c_n(\omega,\mu_R,\qb/\mu_F)\, B_n(\mu_F)
\Big] \,.
\label{f-two}
\end{equation}
The coefficients $c_n$ have the remarkable properties
\begin{equation}
c_n \longrightarrow 1 + \frac{\alpha_s}{\pi} {\cal K}_n \quad {\rm
for}\quad \omega \to 1 \,, \quad \quad
c_n \propto \omega^n \quad {\rm for}\quad \omega \to 0 \,.
\label{limits}
\end{equation}
From (\ref{limits}) it is obvious that, in the real photon limit, the
transition form factor is $\propto 1+\sum B_n$ to LO. To NLO the
sum $\sum B_n$ is slightly resolved due to the running of $\alpha_s$
and evolution. In practice the analyses of $F_{\pi\gamma}$ are
performed with a truncation of the Gegenbauer series.
The simplest analysis assumes $B_n=0$ for $n\ge 4$
\cite{kro96}. A fit to the CLEO data \cite{cleo98} above 
$Q^2_{min}= 2 \gev^2$ then
provides $B_2(1\gev) = - 0.06 \pm 0.03$ to NLO accuracy in the
$\overline{MS}$ scheme. 

If one allows for $B_2$ and $B_4$ in the analysis there is no unique
result for the individual coefficients. Rather there is a strong
linear correlation between both the coefficients; 
only extreme values of $|B_2|$ and $|B_4|$, say above 1 or 2, are
ruled out. A compact way of presenting the result of this fit is to
quote the values of the linear combinations $B_2 + B_4$ and $B_2 - B_4$, 
which have approximately uncorrelated errors: $B_2+B_4 =
- 0.06 \pm 0.08$ and $B_2-B_4 = 0.0 \pm 0.9$ at a scale of $1 \gev$.
This illustrates that, within a leading twist NLO analysis, the CLEO
data on the $\gamma^* \gamma \to\pi$ form factor approximately
fixes only the sum $\sum B_n$ to be close to zero. 

Besides the uncertainties due to the choice of $\mu_F$, $\mu_R$ and
$Q_{min}$ there is another important one in the analysis of the form
factor data that arises from possible power corrections. While our
analysis reveals that logarithmic effects suffice to describe the
residual $Q^2$ dependence of the CLEO data for $Q^2 F_{\pi\gamma}$
above $2 \gev^2$, substantial power corrections cannot be excluded
since it is very difficult to distinguish a power from a logarithmic
behaviour in $Q^2$ with data in the range between 2 and 8 GeV$^2$. 
It is to be emphasized that any estimate of
power corrections is subject to a strong model dependence. Leaving
this out of consideration, one may arrive at misleading results. 

Let me now turn to the case of two virtual photons. From
(\ref{limits}) one sees that for small $\omega$ a Gegenbauer
coefficient $B_n$ is suppressed in $F_{\pi\gamma^*}$ by a power
$\omega^n$. Thus, for small $\omega$, one has
\begin{equation}
F_{\pi\gamma^*}(\qb,\omega) \simeq \frac{\sqrt{2}f_\pi}{3\: \qqb} \left\{
    1 - \frac{\als}{\pi} + \frac{\omega^2}{5}
     \left[ 1-\frac53\frac{\als}{\pi} 
           + \frac{12}{7}\, B_2 
    \left( 1 + \frac{5}{12} \frac{\als}{\pi}\right)\right] \right\}
\label{fpgapprox}
\end{equation} 
The limiting behavior for $\omega\to 0$ has already been given in \cite{agu81}.

\begin{figure}[t]
\begin{center}
\epsfig{file=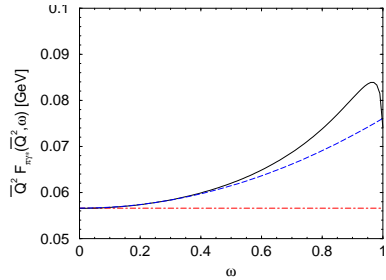,bb=8 12 488 360,width=5cm,clip=}
\end{center}
\caption{{}Comparison of the full result (\ref{fpgvirtual}) for
$\qqb F_{\pi\gamma^*}$ (solid line) with (\ref{fpgapprox}) 
(dashed line) and the $\omega\to 0$ limit (dash-dotted line). 
The form factor is evaluated at $\qb=2 \gev$ for the 
\da{} with $B_2=0.54$, $B_4=-0.40$, $B_6=-0.20$ at a scale of 1 GeV.} 
\label{fig:contour}
\end{figure}
Given the small numerical coefficients in front of $\omega^2$, the
$\omega$ independent term in Eq.~(\ref{fpgapprox}) dominates over a
rather large range of $\omega$. Even at $\omega\simeq 0.6$ the
$\omega^2$ corrections amount to less than $15\%$ if $|B_2|
<0.5$. Thus, for a wide range of $\omega$ the $\gamma^*-\pi$
transition form factor is essentially independent of the pion \da.
To illustrate the quality of the small-$\omega$ approximations we 
compare in Fig.~\ref{fig:contour} the full result
(\ref{fpgvirtual}) for $F_{\pi\gamma^*}$ with the
expression~(\ref{fpgapprox}) for an extreme example of a \da.
The full calculation is in agreement with the CLEO data for
$\omega\to 1$. We see that, although $B_2$ in our example is quite
large and positive, both approximations are indeed very good for
$\omega \lsim 0.6$. Only for $\omega$-values near 1 the form factor is
sensitive to details of the \da. One thus has a parameter-free 
prediction of QCD to leading-twist accuracy.  Any observed
deviation from the limiting behaviour for $\omega\to 0$ beyond what 
can reasonably be ascribed to ${\cal O}(\als^2)$ terms would be an 
unambiguous signal for power corrections. For small $\omega$, 
the limiting behaviour of the form factor has a status
comparable to the famous expression of the cross section ratio $R =
\sigma(e^+ e^-\to {\rm hadrons}) / \sigma(e^+e^-\to \mu^+ \mu^-)$.

The $\gamma^*-\eta$ and $\gamma^*-\eta'$ transition form factors can
be analyzed along the same lines as for the pion. The only
complication is that, to order $\alpha_s$, there is a contribution
from the two-gluon Fock state, its \da{} mixes with the SU(3)-singlet
\da{} under evolution. It has been shown \cite{fel97} that, in the
real photon limit, the CLEO \cite{cleo98} and L3 \cite{acc98} data on
the $\gamma - \eta^({}'^)$ form factors are consistent with
approximately equal \da s for the $\pi$, $\eta$  and $\eta'$ and 
correspondingly vanishing gluon ones.

For small $\omega$ one obtains in analogy to (\ref{fpgapprox}) 
\begin{equation}
F_{P\gamma^*}(\qb,\omega) = 
   \frac{\sqrt{2}\: f_P^{{\rm eff}}}{3\: \qqb}
   \, \bigg[ 1 - \frac{\als}{\pi} \bigg ] 
+ {\cal O}(\omega^2 ,\als^2) \,.
\label{qcd-pre-eta}
\end{equation}
where $f_P^{{\rm eff}}$ are effective, process-dependent decay
constants. Using for instance the quark-flavor mixing scheme 
\cite{fel98}, one finds for the decay constants
$f_{\eta}^{{\rm eff}}=0.98 f_\pi$ and $f_{\eta'}^{{\rm eff}}= 1.62
f_\pi$. At small $\omega$ and large enough $\qqb$ the ratio of the 
$\gamma^*$--$\,\eta,\,\eta'$ form factors 
constitutes an accurate measure of the effective decay constants. 
This can be used for a severe test of the $\eta - \eta'$ mixing scheme. 

In summary: In the real photon limit the transition form factors
essentially provide information on $\sum B_n$ and these sums seem to
be small. 
Data at large $Q^2$ are needed in order to determine the size
of power corrections. 
For $\omega \lsim 0.6$, on the other hand, the
form factors are essentially independent of the \da s. One thus has a
parameter-free QCD prediction which well deserves experimental
verification. Rate estimates for the running $B$-factories reveal that
$F_{\pi\gamma^*}$ should be measurable for $\qqb\lsim 4 \gev^2$ (for a 
luminosity of 30 fb$^{-1}$ per year). 
\section*{Acknowledgments}
It is a pleasure to thank Maria Kienzle and Maneesh Wadhwa for the
well-organized and interesting PHOTON 2001 conference.


\begin{thebibliography}{99}

\bibitem{cleo98} J.\ Gronberg {\em et al.} [CLEO collaboration],
                Phys.\ Rev.\ D {\bf 57}, 33 (1998).
\bibitem{DKV} M.~Diehl, P.~Kroll and C.~Vogt,
              hep-ph/0108220.
\bibitem{agu81} F.\ Del Aguila and M.K.\ Chase,
                Nucl.\ Phys.\ B {\bf 193}, 517 (1981).
\bibitem{bra83} E.\ Braaten, 
                Phys.\ Rev.\ D {\bf 28}, 524 (1983);
		E.P.\ Kadantseva, S.V.\ Mikhailov and A.V.\
                Radyushkin,
                Sov.\ Jour.\ Nucl.\ Phys.\ {\bf 44}, 326 (1986).
\bibitem{lep79} G.P.\ Lepage and S.J.\ Brodsky,
                Phys.\ Rev.\ D {\bf 22}, 2157 (1980).
\bibitem{kro96} P.\ Kroll and M.\ Raulfs,
                Phys.\ Lett.\ B {\bf 387}, 848 (1996).
\bibitem{fel97} T.~Feldmann and P.~Kroll,
		Eur.\ Phys.\ J.\ C {\bf 5}, 327 (1998).
\bibitem{acc98} M.~Acciarri {\it et al.}  [L3 Collaboration],
		Phys.\ Lett.\ B {\bf 418}, 399 (1998).
\bibitem{fel98} Th.\ Feldmann, P.\ Kroll and B.\ Stech,
                Phys.\ Rev.\ D {\bf 58}, 1140006 (1998).

\end{thebibliography}
\end{document}